# Impact of Growth Conditions on the Nature of Magnetism and Magnetotransport of $Sm_{0.55}Sr_{0.45}MnO_3$ Thin Films

Manoj K. Srivastava[1,2], P. K. Siwach[1], A. Kaur[2], and H. K. Singh[1]

[1]National Physical Laboratory, Dr. K. S. Krishnan Road, New Delhi-110012, India
[2]Department of Physics and Astrophysics, University of Delhi, Delhi-110007, India

The impact of growth conditions on the magnetic and electrical transport properties of oriented polycrystalline $Sm_{0.55}Sr_{0.45}MnO_3$ thin films prepared on nearly lattice matching $(La_{0.18}Sr_{0.82})(Al_{0.59}Ta_{0.41})O_3$ (LSAT) (100) single crystal substrates by nebulized spray pyrolysis has been studied. These films were deposited at substrate temperature $T_S = 473$ K and then annealed at $T_A = 1273$ K for 12 hrs in different ambient, viz., air and flowing oxygen. Although the PM-FM transition temperature remains unaffected by the growth conditions and occurs at $T_C = 130$ K in all the films, the nature of magnetism is affected unambiguously. This is well evidenced by the change in (i) the nature of transition, (ii) the magnetic moment, (iii) coercivity, and (iv) remanence; as a function of the synthesis conditions. The electrical transport properties, such as, insulator-metal transition (IMT), temperature coefficient of resistivity (TCR) and magnetoresistance (MR) are also affected by the variations in the growth conditions. The resistivity of all the films shows well defined hysteretic behaviour with respect to the thermal cycling and in the cooling cycle IMT occurs at lower temperature. The nature of the hysteresis and the residual resistivity also shows unambiguous dependence on the growth conditions. Our results demonstrate that oxygen annealing results in (i) reduced magnetic moment, (ii) lower and broad FM-$T_C$, (iii) broadened IMT/lower TCR, and (iv) smaller MR. The observed results have been explained in terms of the growth condition induced variations in the dimensionality and fractions of the competing FMM and AFM-CO phases.

**INTRODUCTION**

Among manganite family, $Sm_{0.55}Sr_{0.45}MnO_3$ is a unique system because of its proximity to the charge order/orbital order (CO/OO) instability and most abrupt insulator metal transition (IMT) [1]-[3]. In general the ground states of this low bandwidth (W) compound are (a) ferromagnetic metallic (FMM) at $0.3 < x \leq 0.52$, and (b) antiferromagnetic insulating (AFMI) for $x > 0.52$. The charge ordering (CO) occurs in the composition range $0.4 \leq x \leq 0.6$ and the corresponding ordering temperature ($T_{CO}$) rises from ~ 140 to 205 K as $x$ increases in the above range[1]-[3]. Colossal magnetoresistance (CMR) is observed at all the compositions that correspond to the FMM ground state in the range $0.3 < x \leq 0.52$ [1]-[3]. In the vicinity of $x \approx 0.45$, a very sharp (first order) transition from paramagnetic insulating (PMI) to the FMM state is observed in $Sm_{0.55}Sr_{0.45}MnO_3$ (SS45). It is also reported that in doped manganites the competition between the FM and AFM/CO state is more clearly seen in the vicinity of half doping where the magnetic phase diagram shows bicriticality. It is well established that the ground state of manganites is dependent on the electronic bandwidth, i.e., the average ionic radius of rare earth/alkaline earth (RE/AE) site cations ($<r_A>$) [1]. The AFM/CO insulating phases are relatively robust in the manganites with a small $<r_A>$ (small W), while the FM metal ones appear in the cases with a relatively larger $<r_A>$. These two phases compete with each other, and the application of a magnetic field can melt the AFM/CO insulating state to the FM metallic state, resulting in the CMR effect.

In the vicinity of the phase boundary, FM and AFM/CO ordered phases are closer in energy, and hence coexist delicately. On the either side of the phase boundary, the nature of transition from the high temperature disordered paramagnetic insulating (PMI) phase to the two ordered phases is akin to the first order [1]. Such abrupt (first order) PMI-FMM transitions are one of the prerequisites of a material having large magnetocaloric effect (MCE) and hence have technological potential [4], [5]. Like other low bandwidth manganites, $Sm_{1-x}Sr_xMnO_3$ has a natural tendency towards phase separation/phase coexistence (PS/PE) [6] that causes evolution of a strong metamagnetic component in the vicinity of half doping. The low W and strong PS result in metamagnetism that render the composition-temperature ($x$-T) phase diagram fragile vis-a-vis external perturbations. Consequently, even mild external stimuli such as electromagnetic field, pressure, lattice strain provided by the substrate, annealing etc. could dramatically modify their physical properties.

Extensive investigation have been made and reported in literature on $Sm_{1-x}Sr_xMnO_3$ having poly- and single- crystalline bulk forms [2]-[8]. However, only limited studies are available on thin films [9]-[12]. Recently Srivastava et al. observed enhancement in PM-FM and IM transition temperatures ($T_C$ and $T_{IM}$) in $Sm_{0.55}Sr_{0.45}MnO_3$ polycrystalline thin films deposited on $LaAlO_3$ (LAO) substrates and attributed this to quenched disorder (QD) caused by the combined effect of oxygen vacancy and compressive strain [12]. However, more studies are required to clarify and distinguish the role of strain and oxygen vacancy in controlling the magnetic and electric transitions in SSMO. In this paper we report the impact of growth conditions on $Sm_{0.55}Sr_{0.45}MnO_3$ polycrystalline thin films deposited on single crystalline (100) $(La_{0.18}Sr_{0.82})(Al_{0.59}Ta_{0.41})O_3$ (LSAT) substrates. The lattice parameter of LSAT (~ 3.865 Å) is only slightly larger than the in plane lattice constants of $Sm_{0.55}Sr_{0.45}MnO_3$ as reported by Tomioka et al [2]. Due to the small lattice mismatch between the substrate and the compound the impact of strain, which is tensile in nature is expected to be small. Thus the impact of growth conditions on the magneto-electrical properties can be easily evaluated.

## EXPERIMENTAL

Polycrystalline $Sm_{0.55}Sr_{0.45}MnO_3$ (SS45) thin films (thickness ~ 100 nm) were deposited on nearly lattice matching LSAT (100) single crystal substrates by ultrasonic nebulized spray pyrolysis [12, 13]. Stoichiometric amounts of 3N pure nitrates of Sm, Sr, and Mn (Sm/Sr/Mn= 0.55/0.45/1) were dissolved in deionized water and the solution was thoroughly homogenized. All the films were deposited at substrate temperature $T_S$ ~ 473 K. To study the role of growth conditions on the magneto electrical properties these as deposited SS45 films were categorized in three sets, one set of films were air annealed and another oxygenated, at the same temperature $T_A$ ~ 1273 K for 12 hrs followed by slow cooling. The third set was annealed in air and then quenched to room temperature (~ 300 K). Hereafter, the films were marked as A, O and Q for air annealed, oxygen annealed and quenched films, respectively. The phase and gross structural analysis was done by X-ray diffraction (XRD) in θ-2θ geometry. The magnetization as a function of temperature and applied magnetic field was measured by PPMS (Quantum Design) at 500 Oe magnetic field applied parallel to the film surface. DC four probe technique has been employed to measure the resistivity of the films.

## RESULTS AND DISCUSSIONS

The highly oriented and polycrystalline nature of the films is brought about the XRD data. This is well evidenced by the dominant appearance of the (0 0 ℓ) (assuming cubic structure) as shown in Fig. 1. The out of plane lattice constant of the air annealed (A), oxygen annealed (O) and quenched films as calculated from the XRD data are 3.819 Å, 3.823 Å and 3.812 Å, respectively. These values are slightly smaller than the c-parameter corresponding to the cubic unit cell, which is c=3.83 Å [2]. Slightly smaller lattice constants in the present case could be attributed to the small tensile strain induced by the LSAT substrate. However, the near constancy of the out of plane lattice parameter of the films prepared under different conditions demonstrates the structural invariance with respect to the annealing conditions.

The temperature dependent zero-field-cooled (ZFC) and field-cooled (FC) magnetization data (M–T) of all the films is shown in Fig. 2. Although the PM-FM transition temperature remains unaffected by the growth conditions and occurs at $T_C \approx 130$ K in all the films, the nature of transition is affected unambiguously. The air annealed-quenched (Q) films show the sharpest PM-FM (that resembles first order) transition, while the oxygen annealed films exhibit the broadest transition. However, in all the films at $T < T_C$, the ZFC and FC branches show strong divergence and at further lower temperatures, the ZFC magnetization shows a sharp drop in all the films. The ZFC magnetization drop is the sharpest in case of the quenched films. The observed low temperature drop in magnetization, coupled with the ZFC-FC divergence is a signature of a metamagnetic state that in case of manganites has been generally identified as the cluster glass (CG) [1], [2], [11]. The sharp drop in the ZFC-MT could be due to cluster freezing that occurs at $T_f \approx 30$ K in the Q-film and at $\approx 25$ K in A- and O- films. The strongest ZFC-FC divergence and sharpest magnetization drop observed in the quenched (Q) films shows that they have the highest cluster glass fraction. The M-H loop of all the films measured at T = 5 K is shown in the inset of Fig. 2. In the air annealed (A) film the coercivity ($H_C$) is found to be $\approx 455$ Oe. The air annealed-quenched films show the largest saturation moment of $M_S \approx 695$ emu/cm$^3$ at $H_S \approx 8.5$ kOe. The $M_S$ of air annealed-slow cooled and oxygen annealed-slow cooled films are $\approx 525$ emu/cm$^3$ (at $H_S \approx 7$ kOe) and 352 emu/cm$^3$ (at $H_S \approx 7$ kOe), respectively. Since the remanence value is directly related to saturation moment hence the reduced remanence ($m_r = M_r/M_S$) were calculated. The air annealed-quenched films (Q) show the largest value of reduced remanence ($m_r=0.47$) while oxygen annealed films (O) show lowest value ($m_r=0.38$). The value of $m_r$ for air annealed-slow cooled films (A) was found to be 0.40. Thus the impact of growth conditions on the nature of magnetism is thus well evidenced by the change in (i) the nature of transition, (ii) the magnetic moment, (iii) coercivity, (iv) remanence, and (v) the saturation moment. The observed variation in the magnetic field and temperature dependent magnetization can be understood in term of the role of oxygen vacancies. In the present case the high temperature annealing in air leads to creation of oxygen vacancies [12], [15]. In the air annealed-slow cooled films these vacancies are expected to ordered, while in the air annealed-quenched films these are expected to be disordered [12], [14, 15]. In manganites, the oxygen vacancies can destabilize the AFM-COI phase both in single crystalline as well as polycrystalline materials quite efficiently [12],[14, 15]. The appreciable larger magnetic moment and sharper PM-FM transition in the two air annealed films thus could be attributed to the melting of AFM-CO clusters due to oxygen vacancies. Thus the smaller value of reduced remanence in the oxygen annealed film could be due to the appearance of AFM-clusters embedded in the FM matrix. Further the sharpening of the PM-FM transition in quenched film could be due to the oxygen vacancy disordering [12]. Since the oxygen stoichiometry is related to the effective hole concentration and even a mild spatial inhomogeneity in the oxygen vacancies could result in spatially varying carrier density that may also act as quenched disorder. The oxygen vacancy annihilation due to annealing can reduce the quenched disorder and affect the magnetic properties as described above.

The temperature dependent resistivity ($\rho$–T) of all the films is shown in Fig. 3. IMT is observed at $T_{IM} \approx 130$ K in the air annealed slowly cooled (A) films and it decreases to $T_{IM} \approx 125$ K in quenched and oxygenated films. Among the three films the Q and O films have the sharpest and broadest IMT, respectively. Throughout the studied temperature range the resistivity of the oxygen annealed films is observed to be higher than the A- and Q- films. This could be due to the appearance of AFM-CO insulating clusters in the FMM matrix. Further, the quenched and oxygenated films show the smallest and largest residual resistivity, which is $\approx 2$m$\Omega$-cm and $\approx 45$m$\Omega$-cm, respectively at 5 K. The nature of the transition also shows a strong correlation with the annealing conditions in addition to the observed variation in resistivity and $T_{IM}$. The resistivity of all the films shows well defined hysteretic behaviour with respect to the thermal cycling and in the cooling cycle IMT occurs at lower temperature. In addition the difference between the resistivity measured during the cooling and warming cycles is more in case of the oxygenated film. The strong hysteresis in the resistivity and large difference in MR could be understood in terms of the cluster glass concept. During the cooling cycle the spin cluster glass behaves like a liquid (disordered) that results in stronger carrier scattering by causing a large resistivity and the lower $T_{IM}$. When cooled below $T_f$ the cluster glass is frozen and the carrier scattering is considerably reduced resulting in lower resistivity and higher $T_{IM}$ during the warming cycle. It is interesting to note that the resistivity of the PM regime is not affected much by the variation in the growth conditions.

The nature of IMT is evaluated from the temperature coefficient of resistivity (TCR) [defined by $TCR\ (\%) = \frac{d}{dT}(ln\rho)x100$] and the same is depicted in Fig. 4. The TCR of the Q-films during the cooling and warming cycles are ≈ 51% and ≈ 44 %, respectively, which decreases to ≈ 31 % and ≈ 26 % in case of the A-films. The oxygenated films have the lowest TCR (and hence broadest IMT) of ≈ 8%. We also measured the low field magnetoresistance (MR) of all the films at H = 3 kOe. The MR measured during the warming cycle is shown in the inset of Fig. 4. The variation of MR as a function of the growth conditions shows nearly the same trend as seen in the TCR. The Q-films possess the largest MR ≈ 30 % that decreases to ≈ 25 % and ≈ 20 % for A- and O- films, respectively. Thus it is clear that electrical transport is also affected unambiguously the growth conditions.

As discussed above the observed variation in the magnetic and electrical transport properties can be understood in term of the role of oxygen vacancies. Our results demonstrate that oxygenation leads to annihilation of oxygen vacancies which favors the growth of the AFM-COI phase at the cost of the FMM and hence results in a broadening of the PMI-FMI transition. Here we must mention that the oxygen vacancy ordering also plays important role. This is well accounted for by the characteristics shown by the quenched (Q) films that show the sharpest PMI-FMM transition, smallest residual resistivity and largest MR. The quenching results in freezing of the randomly distributed oxygen vacancies that transform the correlated quenched disorder into uncorrelated one, which could push the system towards the bicriticality due to the enhanced competition between the FMM and AFM-CO phases. The occurrence of PM-FM and IMT in all the films suggests that the variation in the relative fractions of the FMM and AFM-COI phases is such that the former is always above the percolative threshold. Interestingly, the resistivity and its temperature dependence in the PM-regime remain unaffected by the ordering and disordering of the oxygen vacancies. The hysteresis in the resistivity and concomitant difference in the IMTs could be understood in terms of the spin cluster glass concept. During the cooling cycle the spin cluster glass mimics a liquid like behavior and the carrier scattering by disordered spins is strong. This results in large resistivity and lower $T_{IM}$. When cooled below a temperature corresponding to the cluster freezing temperature ($T_f$), clusters are frozen and in this state the carrier scattering is considerably suppressed leading to the lower resistivity and higher $T_{IM}$ during the warming cycle.

## CONCLUSIONS

In summary, we have synthesized oriented polycrystalline thin films and investigated the effect of the growth conditions and oxygen vacancy ordering/disordering on the nature of magnetism and magnetotransport properties of $Sm_{0.55}Sr_{0.45}MnO_3$. The observed results have been explained in terms of the growth condition induced variations in the dimensionality and fractions of the competing magnetic phases, which in the present case are FMM and AFM/CO-I. Our results show that oxygenation favours the growth of the AFM-COI phase at the cost of the FMM and hence results in a broadening of the PM-FM and IM transitions. The sharpness of the magnetic and MI transitions in the quenched films has been explained in terms of the disordering of the oxygen vacancies.

## ACKNOWLEDGEMENTS

M.K.S. thanks CSIR for research fellowship. Authors are grateful to Prof. R. C. Budhani for support and Dr. V.P.S. Awana for some of the measurements.

**Figure captions-**

Fig.1: XRD pattern of the air annealed (A), oxygen annealed (O), and quenched (Q) $Sm_{0.55}Sr_{0.45}MnO_3$ thin films.

Fig. 2: Temperature dependent ZFC (solid symbols) and FC (open symbols) magnetization (M) of the $Sm_{0.55}Sr_{0.45}MnO_3$ films. The M-H loop of all the films is inserted in the inset.

Fig. 3: Temperature dependent resistivity (ρ) of the $Sm_{0.55}Sr_{0.45}MnO_3$ films measured during the cooling (C) and warming (W) cycles.

Fig. 4: Temperature dependent temperature coefficient of resistivity (TCR) of the $Sm_{0.55}Sr_{0.45}MnO_3$ films measured during the cooling (C) and warming (W) cycles. The inset shows the variation of MR (H@3 kOe) measured during the warming cycle.

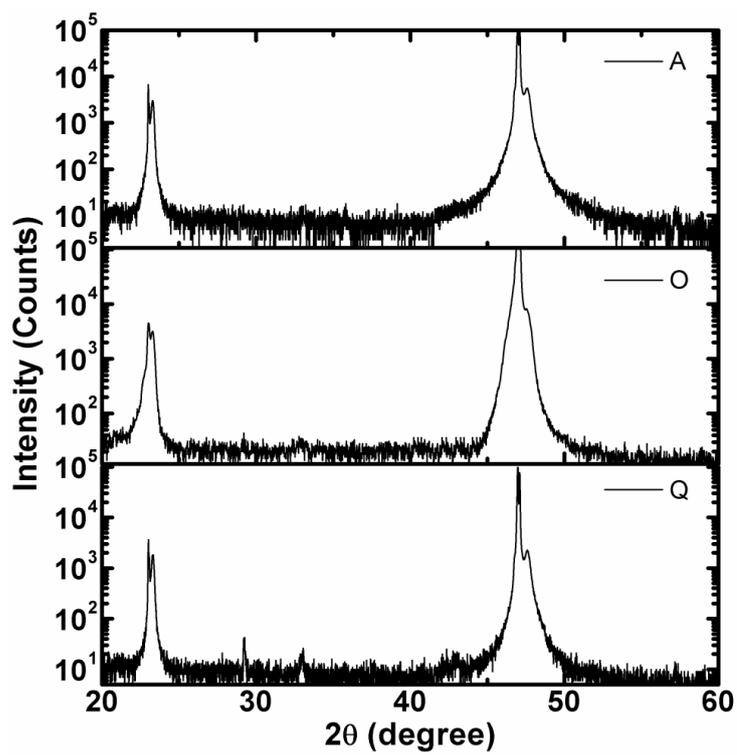

Fig. 1

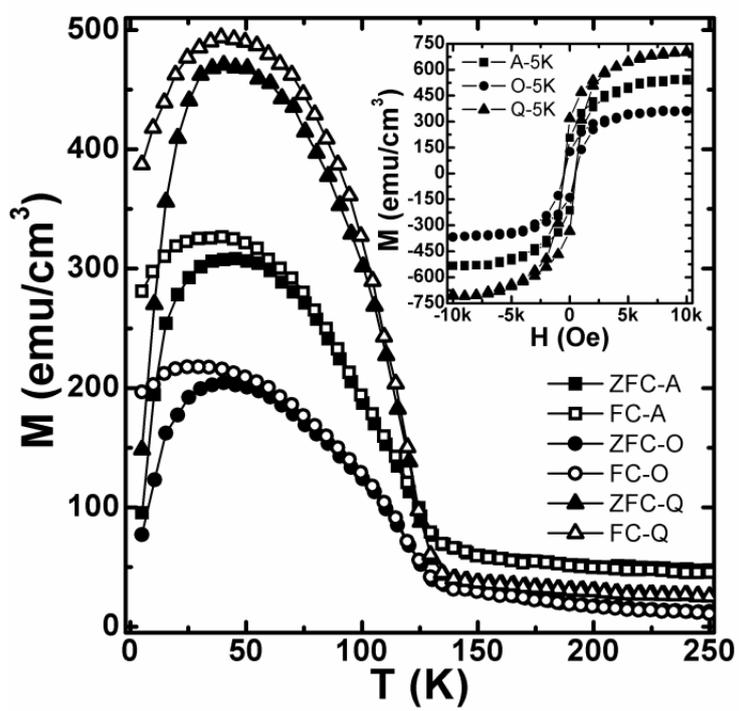

Fig. 2

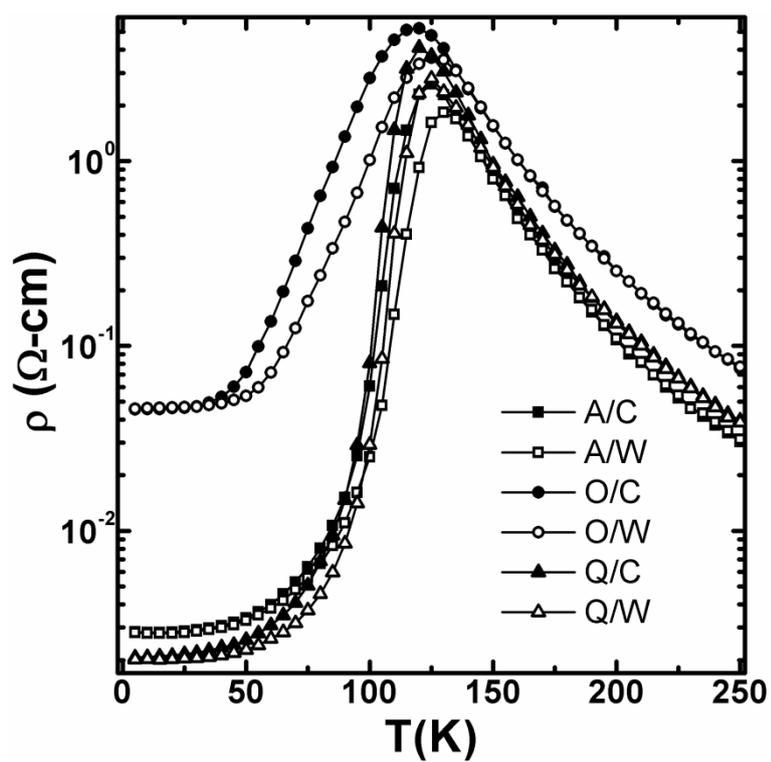

Fig. 3

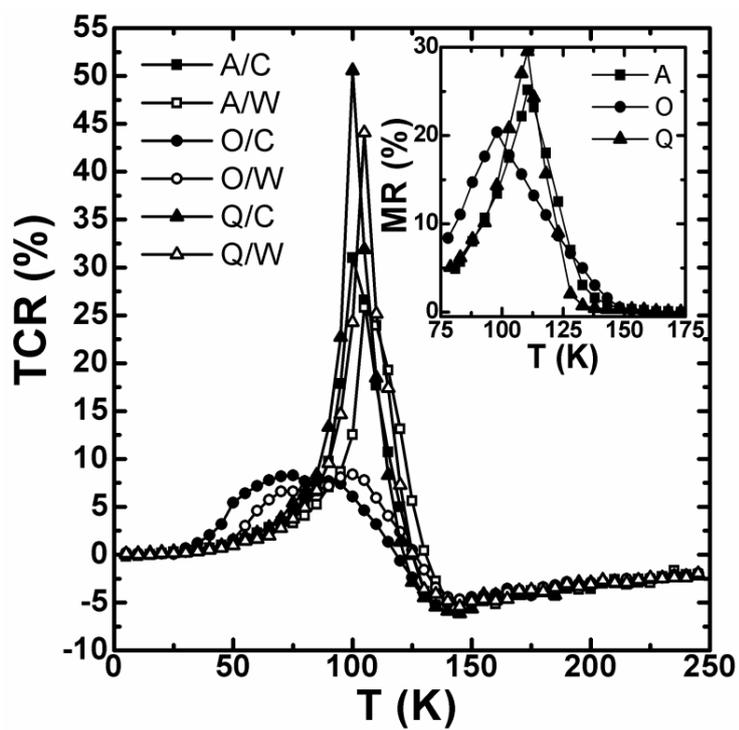

Fig. 4